\documentclass[twocolumn,amsmath,amssymb,10pt,superscriptaddress,nofootinbib]{revtex4}
\pdfoutput=1

\usepackage[latin1]{inputenc}
\usepackage[english]{babel}
\usepackage{amssymb}
\usepackage{amsmath}
\usepackage{amsthm}
\usepackage[]{graphicx}
\usepackage[]{subfigure}
\usepackage{tensor}
\usepackage{color}
\usepackage{cancel}
\usepackage{setspace}
\usepackage{fancyhdr}
\usepackage[bookmarks,linktocpage, colorlinks=true, plainpages = false, citecolor = blue,  linkcolor=blue, urlcolor = blue, filecolor = blue]{hyperref} 

\begin{document}

\allowdisplaybreaks
\begin{titlepage}

\title{Allowable complex metrics in minisuperspace quantum cosmology \vspace{.1in}}

\author{Jean-Luc Lehners \vspace{.1in} \\ {\it Max--Planck--Institute for Gravitational Physics (Albert--Einstein--Institute) \\ 14476 Potsdam, Germany \\ jlehners@aei.mpg.de}}


\begin{abstract}
\vspace{.2in} \noindent 
Kontsevich and Segal (K-S) have proposed a criterion to determine which complex metrics should be allowed, based on the requirement that quantum field theories may consistently be defined on these metrics, and Witten has recently suggested that their proposal should also apply to gravity. We explore this criterion in the context of gravitational path integrals, in simple minisuperspace models, specifically considering de Sitter (dS), no-boundary and Anti-de Sitter (AdS) examples. These simple examples allow us to gain some understanding of the off-shell structure of gravitational path integrals. In all cases, we find that the saddle points of the integral lie right at the edge of the allowable domain of metrics, even when the saddle points are complex or Euclidean. Moreover the Lefschetz thimbles, in particular the steepest descent contours for the lapse integral, are cut off as they intrude into the domain of non-allowable metrics. In the AdS case, the implied restriction on the integration contour is found to have a simple physical interpretation. In the dS case, the lapse integral is forced to become asymptotically Euclidean. We also point out that the K-S criterion provides a reason, in the context of the no-boundary proposal, for why scalar fields would start their evolution at local extrema of their potential. 
\end{abstract}
\maketitle

\end{titlepage}


\section{Introduction and summary} \label{sec:introduction}

Despite the fact that we live in a Lorentzian universe, Euclidean and complex metrics are often used in theoretical physics. For instance, quantum field theories are typically defined on Euclidean 4-space, essentially because free fields can then be described by convergent Gaussian integrals. When gravity is included, a prominent example is provided by black hole metrics in imaginary time, which offer the quickest way of deriving the thermodynamic properties of black holes \cite{Gibbons:1976ue}. It is a little surprising that complex metrics can yield physically sensible results, and the question arises as to which complex metrics ought to be allowed. 

Louko and Sorkin \cite{Louko:1995jw} looked at this question for topology changing transitions in 2 dimensions, and found that only certain kinds of complex deformations make sense. The condition that they employed was to require a scalar field theory to be well defined on the complexified background in question, i.e. that the path integral for a (real) scalar field on a given complex manifold should be convergent. Supporting evidence came from the fact that for such metrics they found that the Gauss-Bonnet integral does not change its value as one deforms the metric within the allowed range, while this integral (which yields a topological invariant) may have a jump in its value for departures from the allowed domain. 

Recently \cite{Kontsevich:2021dmb}, Kontsevich and Segal (K-S) have approached the issue of complex metrics more generally and systematically. What they were chiefly interested in was a determination of complex metrics on which quantum field theories may consistently be defined. They imposed a condition which can be seen as a generalisation of the Louko-Sorkin condition, namely that the path integrals for all $p$-form actions should be convergent. A reason to consider general $p$-form actions is that these admit local covariant stress-energy tensors \cite{Weinberg:1980kq} and hence, by requiring all of these to be well defined, one can ensure that generic local quantum field theories make sense on these complexified backgrounds. We will review their criterion below. Let us highlight that K-S go substantially further, as they propose that on such allowable backgrounds one may construct a Hilbert space and that the theories may be proven to be unitary. An interesting aspect of the K-S condition is that it implies that real Lorentzian metrics reside at the boundary of the domain of allowable metrics, thus their point of view implies that Lorentzian metrics do not sit in the ``middle'' of the space of allowable complex metrics, but rather that Lorentzian metrics can only sensibly be complexified in one direction.

Witten has recently initiated the study of the K-S criterion for dynamical gravity \cite{Witten:2021nzp}, by analysing numerous examples of complex solutions to Einstein's equations. He has shown that the K-S criterion eliminates many pathological complex metrics, such as zero action wormholes, while retaining sensible metrics, such as complexified rotating black holes in asymptotically AdS spacetime. Witten found that all examples that he studied provided support for the validity of the K-S bound, when applied to potential saddle points of the gravitational path integral.

In the present paper we will extend Witten's work by investigating the off-shell structure of gravitational path integrals. The setting we choose consists of the simplest minisuperspace models of quantum gravity. This is because these models offer rather good analytic control, yet they are examples where off-shell configurations play an important role, as one needs to understand the off-shell structure in order to define gravitational path integrals. Crucially, these off-shell configurations consist of complex metrics, since it is only in the space of complex metrics that one can find convergent integration contours for the gravitational path integral. (This is a direct consequence of the fact that the Feynman path integral is an oscillatory, conditionally convergent integral for real metrics.)

We analyse three different settings: transitions between classical boundary conditions, no-boundary path integrals and AdS path integrals. One surprise that we find is that in all three cases the saddle points of the integrals reside at the boundary of the allowable domain of metrics. For classical boundary conditions, this was to be expected, as the saddle points represent Lorentzian geometries. But in the no-boundary case the saddle points are complex, and in the AdS case Euclidean -- nevertheless, the saddle points are in some sense at the edge of the space of metrics. This has the consequence that the steepest descent contours associated with these saddle points necessarily have one end at the saddle points themselves. A direct implication is that one cannot define the contour of integration for the lapse function to run over full thimbles, but only over portions of thimbles ending on saddle points. This may however be a natural ending point, given that the space of allowable metrics has a boundary there. In the AdS case, in particular, we find that the resulting contour runs between two such boundaries and is distinguished by preventing the metric from changing signature, a physically sensible requirement. In the dS and no-boundary cases we find that asymptotically (away from the saddle points) the integration contours become Euclidean. Thus, optimistically,  the present framework may ultimately allow for a consistent definition of Euclidean quantum gravity. Our results certainly provide support for the idea that the Kontsevich-Segal approach has relevance when extended to quantum gravity.

\section{K-S criterion}

As alluded to above, K-S require the action for $p$-form fields (with field strength $F_{j_1 j_2 \cdots j_{p+1}}$) to be well defined, in the sense that a path integral over the kinetic term should be able to converge for all field configurations \cite{Kontsevich:2021dmb,Louko:1995jw,Witten:2021nzp},
\begin{align}
& |e^{\frac{i}{\hbar}S}|<1 \,\,\, \textrm{or} \,\,\, |e^{-\frac{1}{\hbar}S_E}|<1 \,\,\, \textrm{implying} \\  & Re\left[ \sqrt{g} g^{j_1 k_1} \cdots g^{j_{p+1} k_{p+1}} F_{j_1 \cdots j_{p+1}} F_{k_1 \cdots k_{p+1}}\right] > 0\,. \label{KS}
\end{align}
This condition is motivated by the fact that $p$-form fields admit local covariant stress-energy tensors \cite{Weinberg:1980kq}, and as such they can be used to define local quantum field theories. Conversely, restricting to complex manifolds on which all (real) $p$-form actions are convergent means that one has the possibility of defining consistent local quantum field theories on such backgrounds. Locally, one can write the metric in diagonal form
\begin{align}
g_{jk} = \delta_{jk} \lambda_j
\end{align}
where the $\lambda_j$ are complex numbers in general. Then, e.g. in $4$ dimensions, the convergence criterion \eqref{KS}, imposed for $p=0,$ becomes the condition 
\begin{align}
-\pi < Arg(\lambda_1) +Arg(\lambda_2) + Arg(\lambda_3) + Arg(\lambda_4) <\pi\,.
\end{align}
For $p=1,$ there is one inverse metric in addition, which will flip the sign of one of the terms in the relation above. For $p=2,$ two signs would be flipped. Imposing the condition \eqref{KS} for all $p$ then becomes equivalent to the requirement \cite{Kontsevich:2021dmb}
\begin{align}
\Sigma= \sum_{j} |Arg(\lambda_{j})| < \pi\,. \label{bound}
\end{align} 
In words, the sum of the absolute values of the arguments of the metric components must remain below the critical value $\pi.$ Note that this bound is a pointwise criterion, which in the present context will mean that it must be satisfied at all times.

An immediate consequence of this condition is that the Minkowski metric, and real Lorentzian metrics in general, reside right at the boundary of the allowed domain, as they have $\Sigma=\pi.$ This is a reflection of the fact that Lorentzian path integrals are not absolutely convergent, but only conditionally convergent. A further consequence is that spacetimes with more than one time dimension are immediately ruled out. 

K-S were chiefly interested in quantum field theories on non-trivial backgrounds. But it seems natural (and useful) to explore the consequences of this bound when gravity is included and dynamical \cite{Witten:2021nzp}. Minisuperspace models, in which the metric is reduced to a small number of free functions, provide a useful starting point as they are tractable, yet retain the crucial quantum gravitational aspect of allowing one to perform a sum over metrics. Thus we will explore the bound \eqref{bound} in several minisuperspace settings involving both classical and non-classical boundary conditions.

\section{Classical boundary conditions, for a positive cosmological constant} \label{sec:cc}

We will be interested in gravitational path integrals in the presence of a cosmological term $\Lambda$, of the form
\begin{align}
\Psi = \int_{H_0}^{H_1} Dg \, e^{\frac{i}{\hbar}\int d^4x \sqrt{-g} \left[ \frac{R}{2} - \Lambda \right] + c_{0,1} \int d^3y \sqrt{g}K\mid_{H_{0,1}}}
\end{align}
Depending on the boundary conditions and integration contours, these will either represent transition amplitudes between two 3-dimensional hypersurfaces $H_{0,1}$, or wave functions on a given hypersurface $H_1$ (with appropriate conditions on $H_0$, see the discussion below). The boundary conditions are encoded in the coefficients $c_{0,1}:$ for Dirichlet boundary conditions we have $c_0 = - 1, c_1 = +1$ and the boundary terms involve the trace of the extrinsic curvature $K.$ A vanishing value of  $c_{0,1}$ leads to Neumann boundary conditions instead. We will first analyse the case of Dirichlet boundary conditions, for the case where the universe transitions between two specified values of the scale factor. This case will mainly serve to set up notation, but it already highlights some rather general properties.

It is useful to restrict to a closed Robertson-Walker metric written in the form \cite{Halliwell:1988ik}
\begin{align}
ds^2 = - \frac{N^2}{q}dt^2 + q \, d\Omega_3^2\,, \label{metric}
\end{align}
where $N$ is the lapse function, $q(t)$ the square of the scale factor and $d\Omega_3^2$ the metric on a unit 3-sphere of volume $2\pi^2.$ This simple minisuperspace model is described (after integrating by parts) by the action
\begin{align}
S = 2\pi^2\int dt \left[ - \frac{3}{4N}\dot{q}^2 + 3N - N \Lambda q\right]\,, \label{actionmini}
\end{align}
where a dot denotes a derivative w.r.t. $t.$ Note that the action is quadratic in $q.$ The canonical momentum $p$ conjugate to $q$ is given by
\begin{align}
p =  -   \frac{3 \pi^2}{N}  \dot{q}\,.
\end{align}
The equation of motion and constraint respectively read
\begin{align}
\frac{\ddot{q}}{N^2}=\frac{2\Lambda}{3}\,,\quad \frac{\dot{q}^2}{4N^2}= \frac{\Lambda}{3}q-1\,.
\end{align}

\begin{figure}[h]
	\centering
	\includegraphics[width=0.45\textwidth]{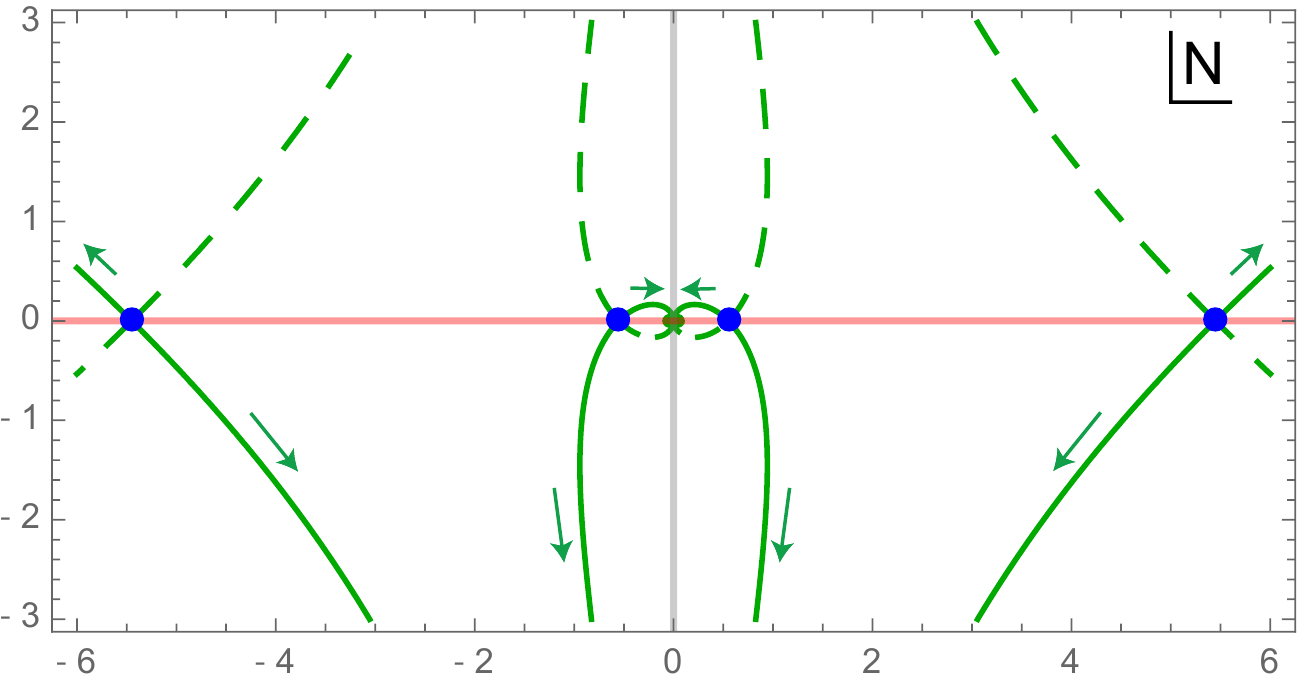}
	\caption{An example with classical boundary conditions. Here we set $\Lambda=3, q_0 = 7, q_1 = 10$ and show the plane of the complexified lapse function $N.$ The saddle points are in blue; steepest descent contours are the solid green lines, while steepest ascent contours are shown by the dashed green lines. The red line indicates Lorentzian metrics, which reside at the boundary of the domain of allowable metrics.}
	\label{Classical}
\end{figure}

As boundary conditions we will impose $q(t=0)\equiv q_0\,, q(t=1) \equiv q_1$, where we have chosen the time coordinate such that the initial and final hypersurfaces reside at $t=0,1$ respectively. The total physical time elapsed between the two hypersurfaces is determined by the lapse $N.$ The solution to the scalar equation of motion, though not necessarily the constraint, is given by
\begin{align}
\bar{q}(t)= \frac{\Lambda}{3}N^2 t^2 + (q_1-q_0-\frac{\Lambda}{3}N^2) t + q_0\,.
\end{align}
Using this solution, the path integral over $q$ may be done by shifting variables to $q=\bar{q}+Q(t),$ where $Q(t)$ is an arbitrary function vanishing at $t=0,1$ (for details, see~\cite{Feldbrugge:2017kzv}). This turns the integral over $q$ into a Gaussian over $Q$ (which merely contributes a prefactor that we will not consider here), leaving us with an ordinary integral over the lapse, 
\begin{align}
& \Psi(q_0,q_1)= \int dN e^{\frac{i}{\hbar}S_c}\,, \\
& \frac{1}{2\pi^2}S_c(N) = \frac{\Lambda^2}{36}N^3 + (3-\frac{\Lambda}{2}(q_0 + q_1))N - \frac{3(q_1-q_0)^2}{4N}\,.
\end{align}
This last integral admits four saddle points, located at 
\begin{align}
N_s = \frac{3}{\Lambda}\left[ \pm \left(\frac{\Lambda}{3}q_1 - 1\right)^{1/2} \pm \left(\frac{\Lambda}{3}q_0 - 1\right)^{1/2}\right] \,.
\end{align}
The locations of the saddle points and their associated steepest descent contours are shown in Fig. \ref{Classical}.

The saddle points at negative $N$ are simply time reverses of those at positive $N.$ The saddle at small positive $N$ describes a universe that expands monotonically from $q_0$ to $q_1,$ while for the saddle at large $N$ the geometry first shrinks to the waist of the de Sitter hyperboloid and then re-expands to the final value $q_1.$ 

In this example the real $N$ line represents Lorentzian metrics, and thus this line represents the boundary of the domain of allowable metrics. The saddle points thus sit right on the boundary, as expected. But this implies that the steepest descent contours (also known as Lefschetz thimbles) are cut at the location of the saddle points. A standard Wick rotation effectively corresponds to a deformation into the lower half plane, thus one would be led to define the path integral only over the portion of the thimbles residing in the lower half plane. Asymptotically these thimbles approach the negative imaginary axis, on which the metrics are Euclidean and $\Sigma=0$. 

Note that it becomes impossible to define a Lorentzian path integral running over the full or half real $N$ line: for asymptotic convergence the integration contour would have to lie in the upper half plane, which is the region of the wrong Wick rotation \cite{Feldbrugge:2017fcc}. This exhibits a tension between the K-S criterion and the Lorentzian viewpoint discussed in \cite{Feldbrugge:2017kzv,Feldbrugge:2017mbc}. Alternatively one could contemplate a contour running between the outer saddle points just below the real $N$ line and passing below the pole at $N=0,$ but then the two thimbles associated with the saddles at small $|N|$ would essentially cancel each other, thus eliminating the possibility of the universe expanding from today until tomorrow -- rather one would predict it to undergo a drastic bounce! In fact it seems most sensible to define the path integral with a contour running either in between the two saddles on the positive $N$ axis (or on the negative axis), or from negative imaginary infinity to one of the saddles (which would render the integral asymptotically Euclidean). The latter would be closest in spirit to the examples we will discuss below. All in all, it might be more reasonable to use different boundary conditions, where the extrinsic curvature is specified \cite{Bousso:1998na,Witten:2018lgb}; this would allow one to include information about the expansion rate of the universe. Within the present minisuperspace ansatz, such conditions however overconstrain the solution for the scale factor, implying that one must generalise the model. We leave this for future work. One context in which a momentum condition has already proven useful is the no-boundary proposal, to which we turn next.

\section{No-boundary proposal}

One of the crucial questions in cosmology is how to determine the initial conditions for the universe. The best studied proposal in this vein is the no-boundary proposal of Hartle and Hawking \cite{Hartle:1983ai}. This is formulated in semi-classical quantum gravity, by restricting the sum over metrics -- a context that is therefore ideally suited to test the K-S criterion. 

The main idea of the no-boundary proposal is that the wave function of the universe should be given by a sum over metrics that have the present $3$-dimensional hypersurface as their only boundary, and for which the  geometries are rounded off in the past. Given that there is not supposed to be an initial boundary, in order to implement the proposal we simply do not put a boundary term on the first hypersurface ($c_0=0$). Consistency of the variational problem then forces us to impose a Neumann condition there \cite{Louko:1988bk}. More specifically, obtaining regular geometries as the universe shrinks to zero size requires the momentum condition (see for instance \cite{DiTucci:2019bui})
\begin{align}
p\mid_{t=0} \equiv p_0 = -6\pi^2 i \,. \label{momcond}
\end{align} 
The sign on the right hand side is chosen such that small tensor perturbations acquire a Gaussian distribution, rather than an inverse Gaussian one. The solution for the scale factor respecting this condition and reaching a final value $q_1$ at $t=1$ is given by
\begin{align}
\bar{q}_{ND}(t)=\frac{\Lambda}{3}N^2 (t^2 -1) +2 N i (t -1) + q_1\,. \label{metricND}
\end{align}
Once again the path integral over $q$ can be done by shifting variables, and results in the wave function being given simply by a lapse integral,
\begin{align}
\Psi(p_0,q_1) & = \int dN e^{i(S_0)/\hbar}\,, \label{lapsepartition}\\
\frac{1}{2\pi^2} S_0(N)  &= \frac{\Lambda^2}{9}N^3+i\Lambda N^2 - \Lambda q_1 N - 3 q_1 i \,. \label{sphereaction}
\end{align}
This time there are only two saddle points, which are complex valued and represent geometries that are time reverses of each other,
\begin{align} 
    N_\pm & = \frac{3}{\Lambda}\left(-i \mp \sqrt{\frac{\Lambda}{3}q_1-1} \right)\,, \label{S3saddles} \\
    S_0(N_\pm)     &= \frac{12\pi^2}{\Lambda} \left[-i \pm \left(\frac{\Lambda}{3}q_1-1 \right)^{3/2}\right]\,. \label{saddleactionN}
\end{align}

\begin{figure}[h]
	\centering
	\includegraphics[width=0.25\textwidth]{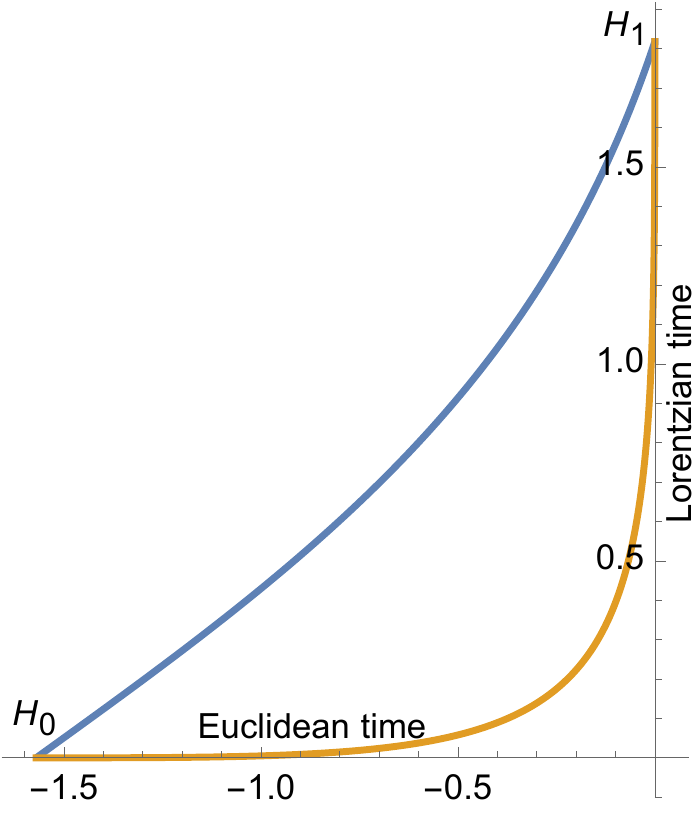} \hspace{0.2cm}
	\caption{A no-boundary saddle point solution, shown in two representations differing by the path taking in the complexified time plane between the South Pole at $H_0$ and the final hypersurface $H_1$. Here we have chosen $\Lambda=3, q_1 =10$ implying that $N_+=3-i.$}
	\label{timecontours}
\end{figure}

Before investigating the status of off-shell geometries, we should first analyse the saddle points themselves. The minisuperspace path integral gives these to us in a different form than the ``standard' Hawking instanton, which consists of half of a (Euclidean) 4-sphere glued to half of a (Lorentzian) de Sitter hyperboloid. This can be viewed as a gluing of a solution with imaginary lapse onto one with real lapse. Here, the saddle point solution we obtain has constant complex lapse, and may be viewed as a kind of shortcut from the ``South Pole'' of the instanton (at $t=0$ where $q=0$) to the final hypersurface (where $t=1$ and $q=q_1$). To compare these representations, it is useful to transform to Euclidean time by defining $i\frac{N}{\sqrt{q}}dt \equiv dT,$ then we get 
\begin{align}
\sqrt{\frac{\Lambda}{3}}\, T(t) = 2i \operatorname{arsinh}\left( \sqrt{\frac{\Lambda Nt}{6i}}\right) - \frac{\pi}{2} \,.
\end{align} 
In Fig. \ref{timecontours} we plot the resulting path in the complex $T$ plane for the transformed constant $N$ solution (in blue). We also plot a different path, with the same end points but an intermediate evolution that is closer to the ``Hawking'' contour, obtained by specifying $T=\theta(t)$ with the choice 
\begin{align}
\theta(t)=-\frac{\pi}{2}(1-t)^n + T(1)\, t^n\,, \quad 0 \leq t \leq 1\,. \label{modtime}
\end{align}
In the plot we chose $n=3$ (orange curve). 

\begin{figure}[h]
	\centering
	\includegraphics[width=0.35\textwidth]{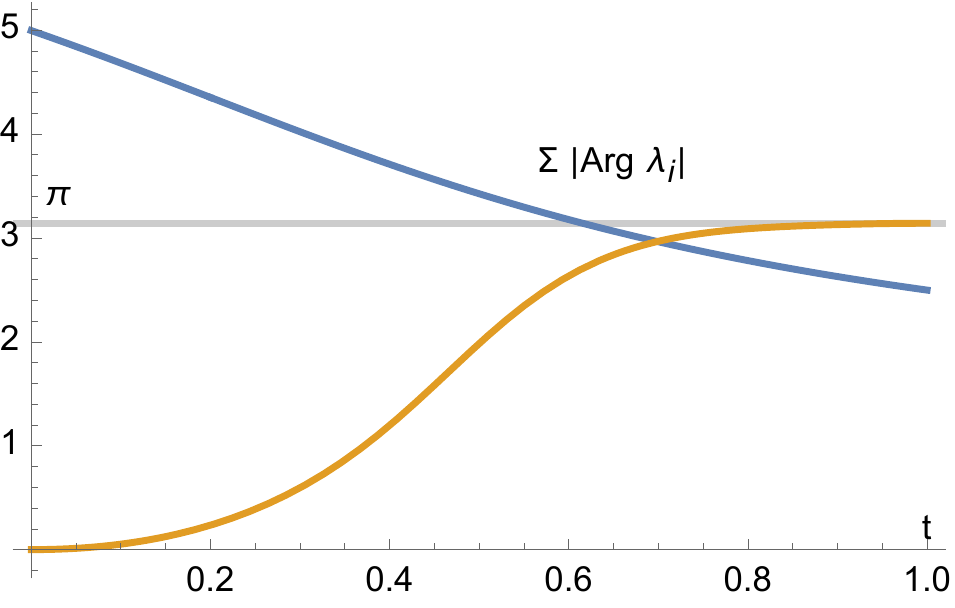}
	\caption{For the two representations shown in Fig. \ref{timecontours}, the K-S criterion behaves very differently: the constant lapse version (in blue) violates the criterion, while the modified path \eqref{modtime} (in orange) is seen to satisfy it everywhere. The grey line indicates the K-S bound, which must not be surpassed.}
	\label{sumarguments}
\end{figure}

The reason for showing this alternate path becomes apparent once we take a look at the associated K-S bound~\eqref{bound}. The values of $\Sigma(t)$ for both paths are shown in Fig.~\ref{sumarguments}. As one can see from this figure, the minisuperspace saddle point solution with constant lapse violates the K-S criterion rather drastically at small $t.$ However, if we deform the time contour to one that more closely follows a Euclidean evolution first, and then approaches a Lorentzian evolution, then the K-S criterion is satisfied at all times (this point was first made in  \cite{Witten:2021nzp}). But really the two representations should be seen as the same solution, since the path in the complexified time plane can be modified according to Cauchy's theorem, as long as the end points remain fixed and no singularities are encountered (see also \cite{Hertog:2011ky}). In particular, both versions lead to the same action and thus the same wave function. We will take the stance that we deem any metric to be allowable if it can be transformed in a similar fashion to a metric that satisfies the K-S criterion. 

The case that we just analysed contained no singularities. For completeness let us point out that when anisotropies are included, no-boundary instantons are known to develop singularities, so that the ability to transform the time contour may be restricted \cite{Bramberger:2017rbv}. It would be interesting to explore whether this might lead to a bound on the anisotropies. We leave this question for future work.

We are now in a position to analyse the off-shell geometries encountered in the path integral, implying that we are interested in the value of 
\begin{align}
\Sigma(t) = |Arg\left( -\frac{N^2}{q(t)}\right)| +  3|Arg\left(q(t)\right)|\,.
\end{align} 
The preceding discussion implies that in general it will be rather difficult to conclusively decide whether a given geometry is allowable or not, as one must in general know all of its possible representations. In other words, we always have the freedom to make similar changes to the complexified time path as the one discussed above in order to reduce $\Sigma(t)$, and in general this is a complicated optimisation problem. That said, it turns out that the present case is actually tractable: just as for the constant lapse saddle point (blue curve in Fig.~\ref{sumarguments}), one may convince oneself that for the geometries specified by \eqref{metricND} the strongest violation of the K-S inequality occurs for small $t.$ Thus we may simply analyse the K-S criterion at $t=0,$ where the value of $q$ is fixed to be $q(0)= q_1 - \frac{\Lambda}{3}N^2 - 2Ni,$ and this value will be \emph{independent} of the time contour chosen. Thus we obtain a lower bound $\Sigma \geq 3|Arg\left(q(0)\right)|,$ and if this bound is violated then we know for sure that the geometry will not be allowable, because even a clever time contour cannot lower the value of $\Sigma.$ (Numerical exploration of several examples has shown that ``good'' time contours, analogous to \eqref{modtime}, can indeed be found for cases where $3|Arg\left(q(0)\right)|$ remains below $\pi$.)

\begin{figure}[h]
	\centering
	\includegraphics[width=0.45\textwidth]{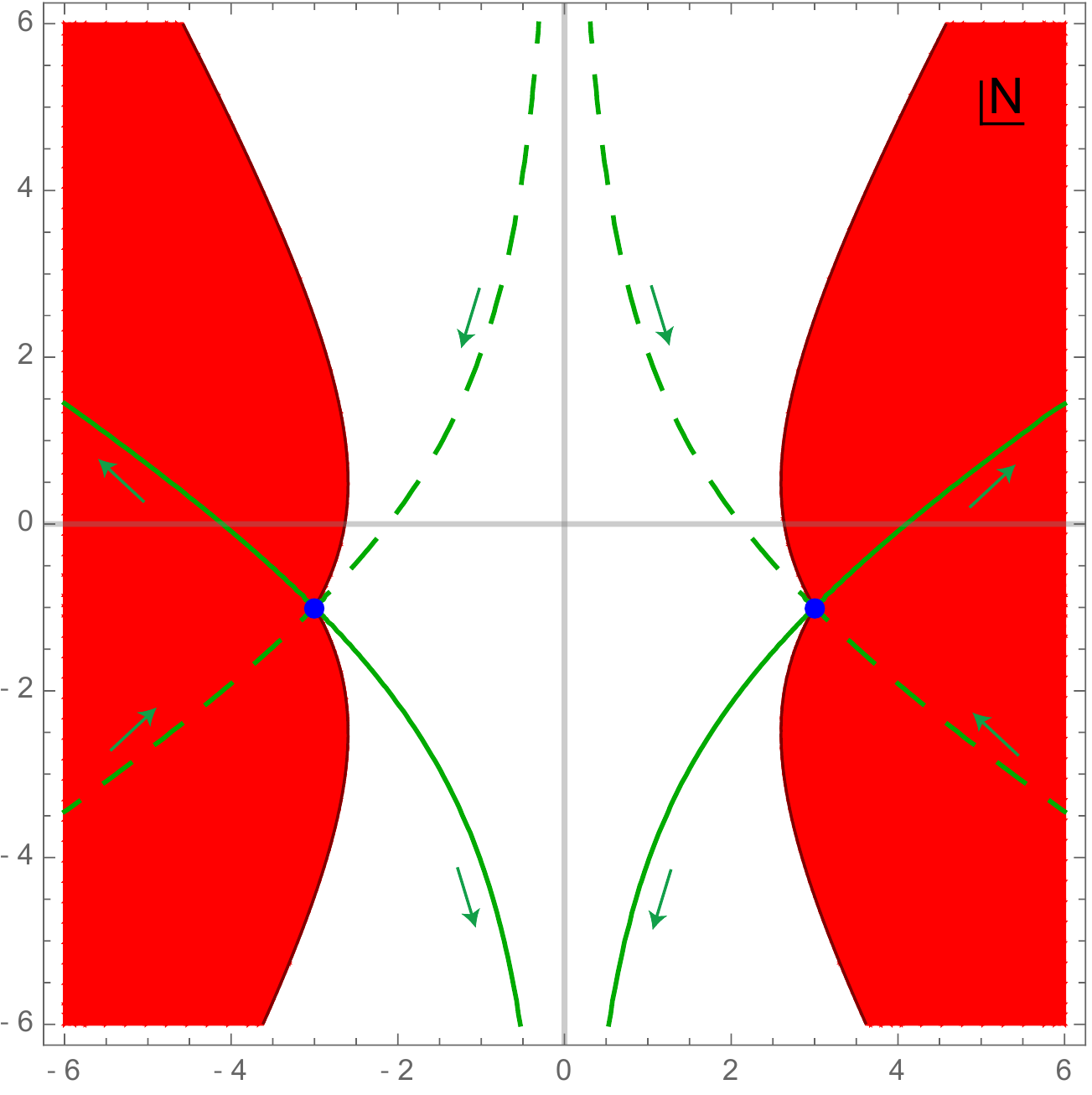}
	\caption{Allowable metrics in the no-boundary proposal, shown in the complex lapse plane. The allowable domain is in white, while the eliminated metrics are in red. Saddle points are marked by blue dots; solid green lines indicate steepest descent contours, dashed green lines steepest ascent contours and arrows indicate downwards flow. Here we used $\Lambda=3, q_1 = 10.$}
	\label{dS}
\end{figure}

Analytically we may already obtain some understanding of the regions where metrics are disallowed. It is useful to first analyse the limit of large absolute value of the lapse. Then we obtain the approximation
\begin{align}
\Sigma(t=0) & \geq  3|Arg(q_1-\frac{\Lambda}{3}N^2 -2Ni)| \\ & \approx 3|Arg(-N^2)| \qquad (\textrm{large }|N|)\,,
\end{align}
where we have used the fact that $\Lambda >0.$ Thus we immediately find that asymptotically the wedges $-\frac{\pi}{3} < Arg(N) < \frac{\pi}{3}$ and $\frac{2\pi}{3} < Arg(N) < \frac{4\pi}{3}$ are disallowed. In particular, the line of real $N$ values is ruled out asymptotically. We can also analyse what happens near the saddle points by writing $N=N_\pm + \Delta,$ and work to linear order in $\Delta.$ There we have
\begin{align}
q(0) \approx  \pm 2 \Delta (\frac{\Lambda}{3}q_1-1)^{1/2}\,.
\end{align} 
We will assume that $q_1>\frac{3}{\Lambda}.$ For the saddle point on the left (3rd quadrant) we thus find the condition $3|Arg(\Delta)|<\pi,$ implying that the allowed wedge is $-\frac{\pi}{3} < Arg(\Delta) < \frac{\pi}{3}.$ For the saddle on the right (4th quadrant) we likewise obtain
$\frac{2\pi}{3} < Arg(\Delta) < \frac{4\pi}{3}.$ Thus we find that, even though the saddle points represent complex geometries, they are surrounded by both allowed and disallowed regions, and reside right at the edge of the allowed domain.

In Fig. \ref{dS} a numerical example of the (dis-)allowed domains is shown. This confirms the expectations we have just developed. The allowable domain resides in between the two saddles, and encompasses the Euclidean axis. The thimbles are cut off at the location of the saddles, and the ``halves'' that remain run off to minus imaginary infinity. This leaves essentially just one possible way of defining the contour of integration for the lapse, if we take into account the fact that the no-boundary wave function is thought to be real \cite{Hartle:1983ai,Lehners:2021jmv}. Namely, one has to sum the two half-thimbles running up from minus imaginary infinity to the saddles. These two contributions are complex conjugates of each other (because $N_+^*=-N_-$), resulting in a real wave function. Note that it does not make sense to take the contour to run over the negative imaginary half-axis, as with the momentum condition \eqref{momcond} there is no pole at $N=0$ and thus this point does not represent a natural end point for the contour. Rather, one has to integrate from negative imaginary infinity, following the thimbles right to the edge of the allowable domain.

\section{AdS path integrals}

We can perform an analogous analysis when the cosmological constant is negative. In this case we expect the saddle points to be given by Anti-de Sitter solutions. There is a close analogy between path integrals with positive and negative $\Lambda,$ relating the no-boundary wave function to the canonical partition function in AdS spacetime \cite{DiTucci:2020weq,Lehners:2021jmv}. If the AdS/CFT conjecture \cite{Maldacena:1997re} is to hold, then we expect the gravitational path integral with negative $\Lambda$ to be well defined \cite{Caputa:2018asc}. This provides a good reason to investigate what the K-S criterion implies in this context.

The formulae from the previous section can be transferred directly, though with $\Lambda<0$ the saddle points become purely imaginary. They correspond to portions of Euclidean Anti-de Sitter space that cap off smoothly at $t=0.$ $N_-$ is the dominant saddle, while $N_+$ corresponds to a saddle containing a second zero in the scale factor, and may thus be expected to be singular upon inclusion of perturbations \cite{DiTucci:2020weq}. 

The strongest constraint on allowable metrics once again is found to come from $t=0.$ We can understand several features of the allowable domain of metrics analytically.
At large $|N|,$ in particular, we can approximate $\Sigma$ by
\begin{align}
\Sigma(t=0) & \geq  3|Arg(q_1-\frac{\Lambda}{3}N^2 -2Ni)| \\ & \approx 3|Arg(+N^2)| \,.
\end{align}
Thus at large $|N|,$ only the wedges $-\frac{\pi}{6} < Arg(N) < \frac{\pi}{6}$ and $\frac{5\pi}{6} < Arg(N) < \frac{7\pi}{6}$ belong to the allowable set of metrics, and the asymptotic Euclidean directions are eliminated. Further note that near the saddle points $N=N_{\pm}+\Delta,$ we again find 
\begin{align}
q(0) \approx -2 \Delta \left(\frac{\Lambda}{3} N_{\pm} + i\right) = \pm 2 \Delta (\frac{\Lambda}{3}q_1-1)^{1/2}\,. \label{qzero}
\end{align} 
Thus for $N$ just above the saddle $N_+$ in the upper half plane, or just below $N_-$ in the lower half plane, $q(0)$ becomes negative and $\Sigma \geq 3\pi.$ This implies that once again the saddle points reside directly at the boundary of the allowable domain, even though the saddles are purely Euclidean.

\begin{figure}[h]
	\centering
	\includegraphics[width=0.45\textwidth]{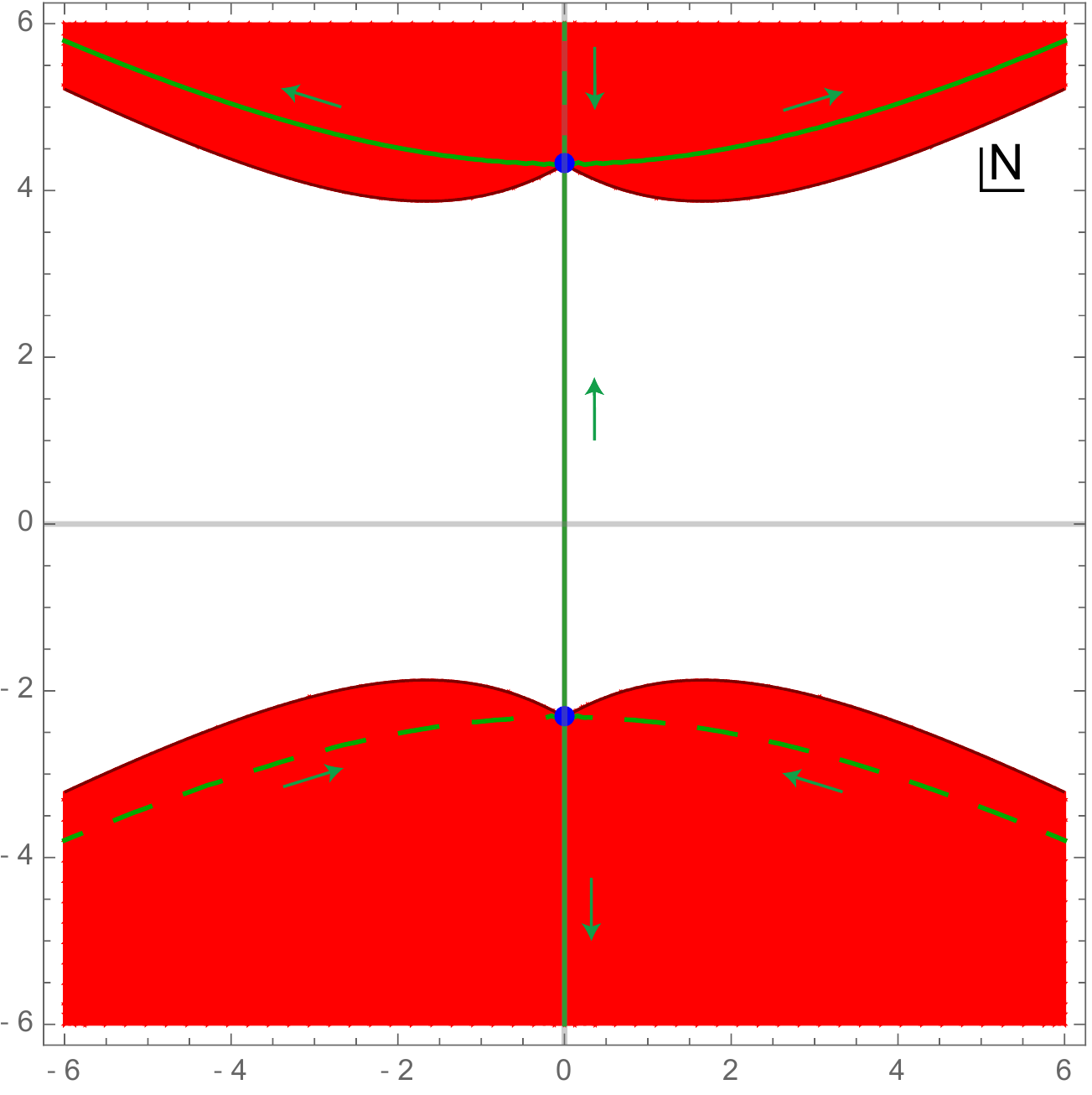}
	\caption{Allowable metrics for the AdS minisuperspace path integral, shown in the complex lapse plane. Conventions are the same as for Fig. \ref{dS}, only that here $\Lambda=-3.$}
	\label{AdS}
\end{figure}

A graph of the numerically determined allowable domain is shown in Fig. \ref{AdS}. This confirms the expectations from the analytic approximations just presented. The thimbles are cut off at the locations of the saddles, and this time only the portion of a thimble linking the two saddles remains. One would thus define the path integral by an integration over precisely this portion of the Euclidean lapse axis. From \eqref{metricND} and \eqref{qzero} it is interesting to note that this portion of the Euclidean axis is distinguished by the metrics having an everywhere positive scale factor squared. Hence, although we cannot define the integration to run over the entire Euclidean axis  (along which the lapse integral in any case would not converge), the integration contour is nevertheless in line with the sound physical requirement that signature change should not occur.  

\section{Further comments}

We have seen that the K-S criterion restricts the integration contours for the lapse integral, and that it thus restricts the possible dynamical evolutions of the universe. This may be a welcome feature, since it may ultimately provide reliable predictions from quantum gravity. Though this is a long term goal, let us remark on a possible consequence in the context of the no-boundary proposal. There, if one adds a scalar field, one finds that at generic locations in the scalar potential, finding a regular no-boundary solution requires the scalar field to be complex valued \cite{Lyons:1992ua}. This runs against the assumptions inherent in the criterion for allowable metrics \cite{Louko:1995jw,Kontsevich:2021dmb}. If one now insists on the scalar field taking real values, then the only solutions that remain are those where the scalar sits at an extremum of the potential. In the no-boundary setting, small values of the potential come out as favoured \cite{Hartle:2008ng}, so that it would be implied that scalars would preferentially sit at a local minimum. In the context of string theory, this might help explain why physical constants are not observed to change over space and time. Moreover, if one scalar starts out in an ``excited'' state, it would be found at the top of a local maximum of the potential, which might be able to explain the initial conditions for an inflationary phase. 

From what we have seen, it is clear that the restriction to allowable metrics can be incorporated rather naturally in the path integral approach to (semi-classical) quantum gravity, while it is at present difficult to see how one would implement such a criterion at the level of the Wheeler-DeWitt equation. This is a question for future research. 

In closing, let us note that the restriction to allowable metrics (in the Kontsevich-Segal sense) may lead to a refinement of the old paradigm of Euclidean quantum gravity: our results demonstrate that, at least in simple settings, one is led to choose integration contours that asymptotically correspond to Euclidean metrics. However, instead of summing over purely Euclidean metrics, one should rather follow the Lefschetz thimbles and interpolate between this Euclidean infinity, where $\Sigma$ vanishes, and the boundary of allowable metrics, where $\Sigma$ reaches $\pi$. There, at the boundary, is where the interesting solutions lie.

\vspace{0.2in}
\acknowledgments

I would like to thank Axel Kleinschmidt and Edward Witten for stimulating discussions and correspondence.
I gratefully acknowledge the support of the European Research Council in the form of the ERC Consolidator Grant CoG 772295 ``Qosmology''.

\bibliographystyle{utphys}
\bibliography{ComplexMetrics}

\end{document}